\documentclass[%
reprint,
superscriptaddress,
amsmath,amssymb,amsthm,mathrsfs,
aps,
prl,
]{revtex4-1}

\usepackage{graphicx}
\usepackage{dcolumn}
\usepackage{bm}
\usepackage[usenames]{color}
\usepackage[ 
margin=0.75in,
]{geometry}
\usepackage[normalem]{ulem}

\def\Tb{$T_B$}

\def\fSPM{$f_{\mathrm{SPM}}$}
\def\muSR{$\mu$SR}

\def\feo{Fe$_3$O$_4$}
\newcommand{\ra}[1]{\renewcommand{\arraystretch}{#1}}

\begin{document}
	
\title{
Superparamagnetic dynamics and blocking transition in Fe$_3$O$_4$ nanoparticles probed by vibrating sample magnetometry and muon spin relaxation
}
	
	\author{Benjamin A. Frandsen}
	\email{benfrandsen@byu.edu}
	\affiliation{ %
	Department of Physics and Astronomy, Brigham Young University, Provo, Utah 84602, USA.
	} %

	\author{Charlotte Read}
	\affiliation{ %
		Department of Physics and Astronomy, Brigham Young University, Provo, Utah 84602, USA.
	} %
	
	\author{Jade Stevens}
	\affiliation{ %
		Department of Physics and Astronomy, Brigham Young University, Provo, Utah 84602, USA.
	} %

	\author{Colby Walker}
	\affiliation{ %
		Department of Physics and Astronomy, Brigham Young University, Provo, Utah 84602, USA.
	} %

	\author{Mason Christiansen}
\affiliation{ %
	Department of Physics and Astronomy, Brigham Young University, Provo, Utah 84602, USA.
} %

\author{Roger G. Harrison}
\affiliation{ %
	Department of Chemistry and Biochemistry, Brigham Young University, Provo, Utah 84602, USA.
} %

\author{Karine Chesnel}
\email{kchesnel@byu.edu}
\affiliation{ %
	Department of Physics and Astronomy, Brigham Young University, Provo, Utah 84602, USA.
} %

\begin{abstract}
The magnetic properties of \feo\ nanoparticle assemblies have been investigated in detail through a combination of vibrating sample magnetometry and muon spin relaxation (\muSR) techniques. Two samples with average particle sizes of 5~nm and 20~nm, respectively, were studied. For both samples, the magnetometry and \muSR\ results exhibit clear signatures of the superparmagnetic state at high temperature and the magnetically blocked state at low temperature. The \muSR\ data demonstrate that the transition from the superparamagnetic to the blocked state occurs gradually throughout the sample volume over a broad temperature range due to the finite particle size distribution of each sample. The transition occurs between approximately 3~K and 45~K for the 5~nm sample and 150~K and 300~K for the 20~nm sample. The magnetometry and \muSR\ data are further analyzed to yield estimates of microscopic magnetic parameters including the nanoparticle spin-flip activation energy $E_A$, magnetic anisotropy $K$, and intrinsic nanoparticle spin reversal attempt time $\tau_0$. These results highlight the complementary information about magnetic nanoparticles that can be obtained by bulk magnetic probes such as magnetometry and local magnetic probes such as \muSR.
\end{abstract}
	
\maketitle

\section{Introduction}	
Magnetic nanoparticles have stimulated significant interest over the past several decades both as a platform for fundamental studies of magnetism on the nanoscale as well as a route to novel technologies in sectors ranging from information technology to medicine~\cite{pankh;jpd03,de;am08,frey;csr09,majet;acsn11}. One of the interesting phenomena associated with magnetic nanoparticles is that of superparamagnetism, whereby ferro- or ferrimagnetic nanoparticles small enough to be in the mono-domain limit exhibit strong thermal fluctuations of the magnetization. Referring to the net magnetic moment on each nanoparticle as a nanospin, we can describe superparamagnetism as the dynamical flipping of nanospins under the influence of temperature, similar to the random flipping of electronic magnetic moments in a conventional paramagnet. When cooled to sufficiently low temperatures, the nanospins freeze, thereby entering the blocked state. This blocking transition is a complex process that depends on the characteristics of the individual nanoparticles in question and can also be strongly influenced by any inter-particle interactions present in the nanoparticle assembly~\cite{goya;jap03,lima;jap06,varon;sr13}.

Although much experimental and theoretical work has been performed to understand superparamagnetism and the blocking transition~\cite{majet;jpd06,roca;nano06,yang;jpcb05}, many details relating to these behaviors remain unclear. Among others, these behaviors include the precise manner in which the the blocking transition proceeds throughout the sample, and how this affects experimental data~\cite{lives;sr18}; the influence of the structure and morphology across varying length scales~\cite{varon;sr13,klomp;ieeem20}; and even the estimation of microscopic parameters such as the nanoparticle spin flip activation energy and fluctuation time~\cite{perig;aprev15}. There is increasing recognition that a more complete understanding is most likely to be obtained by attacking the problem with a variety of complementary techniques, such as magnetometry~\cite{roca;nano06}, M\"ossbauer spectroscopy~\cite{gabba;jmmm15}, electronic microscopy~\cite{yamam;prb02} and holography~\cite{yamam;apl08}, neutron scattering~\cite{mishr;jpcm15}, x-ray magnetic spectroscopy~\cite{cai;jap14}, or x-ray magnetic scattering~\cite{chesn;mag18}. In this context, probes of local magnetism---i.e. methods that are sensitive to magnetism on length scales of several interatomic spacings---provide important information not easily obtained from bulk-averaged probes such as conventional magnetization measurements. Muon spin relaxation/rotation (\muSR) is one such local magnetic probe that has heretofore seldom been applied to the topic of superparamagnetism. This technique can provide a wealth of information, such as the volume fraction of competing magnetic states and the nature of magnetic fluctuations on time scales of 10$^{-12}$ to 10$^{-4}$~s (highly complementary to slower probes such as nuclear magnetic resonance or ac/dc magnetic susceptibility measurements and faster probes such as neutron scattering)~\cite{uemur;ms99,blund;cp10,yaoua;b;msr11}.

In this work, we combine \muSR\ experiments with vibrating sample magnetometry (VSM) to investigate the superparamagnetic behavior and blocking transition in \feo\ nanoparticle assemblies. Two samples with average particle sizes of approximately 5~nm and 20~nm were studied. As established in previous studies~\cite{chesn;jpconfs14,klomp;ieeem20}, we find that the blocking temperature \Tb\ increases significantly as the particle size is increased from 5~nm to 20~nm. The magnetometry results provide insight into the drastic dependence of the blocking temperature on particle size and on how the particle size distribution affects the extent of the blocking transition, but quantitative information about the size distribution dependence of the blocking transition requires complex modeling~\cite{lives;sr18}. Even without such modeling, however, we are still able to obtain average information from the magnetometry data, including various estimates of the average blocking temperature. We then use the \muSR\ data for more quantitative analysis of the blocking transition. We establish that the blocking transition occurs gradually throughout the sample volume over a broad temperature interval due to the finite size distribution of the particles. More importantly, we extract crucial microscopic magnetic parameters directly from the temperature-dependent \muSR\ relaxation rate, including the nanoparticle spin flip activation energy, the magnetic anisotropy, and the nanoparticle spin fluctuation time (i.e., the magnetic reversal attempt time). Our study shows that, complementary to magnetometry, the \muSR\ technique provides an accurate and independent means of probing crucial properties of magnetic NP assemblies.

\section{Experimental Methods}	
\feo\ NPs with average diameters of approximately 5~nm were synthesized using an organic solution method detailed in Ref.~\onlinecite{altav;am05}, and made of \feo\ particles surrounded by a ligand shell of oleic acid molecules. To purify the NPs, they were dissolved in toluene, precipitated from ethanol, and extracted after centrifugation at 5000 revolutions per minute for approximately 15 minutes. The end material was a black paste containing a dense assembly of the NPs. \feo\ NPs with average diameters of approximately 20~nm were purchased from Cytodiagnostics Inc in Ontario, Canada. These NPs came dissolved in toluene and were similarly encased in an oleic acid ligand shell and were shipped in solution with toluene. The toluene was evaporated by bubbling nitrogen gas through the sample, leaving a thick residue consisting of the NPs. Bulk \feo\ powder for comparison with the NPs was purchased from Alfa Aesar.

The transmission electron microscopy (TEM) images were collected on a ThermoFisher Scientific Tecnai F20 UT operating at 200~kV. For TEM imaging, the NPs were deposited on very thin carbon membranes by dropping solutions of the NPs on the membranes and letting the solvents, either toluene or chloroform, evaporate. For each NP sample, the concentration of the solution was finely adjusted so as to achieve a monolayer or less of NPs on the membrane, thus allowing individual NPs to be observed. 

The vibrating sample magnetometry (VSM) data were collected on a Quantum Design Physical Properties Measurement System (PPMS) that includes a superconducting magnet, generating a ramping field up to 9 T, and a cryogenic sample holder using liquid helium. For the VSM measurements, a small amount of the dried NP material was inserted into capsules of approximate volume 1 mm$^3$. The material was tightly compacted inside each capsule, which was then placed inside a cylindrical brass sample holder to be inserted in the VSM coilset. Field cooling (FC) and zero-field cooling (ZFC) measurements were carried out under a field of 100~Oe (10~mT) at a warming speed of 1~K/min over a temperature range of 10~K – 400~K. Data points were collected in intervals of 1~s.

The \muSR\ experiments were performed at TRIUMF in Vancouver, Canada on the LAMPF (Los Alamos Meson Physics Facility) spectrometer. The \muSR\ technique involves implanting spin-polarized positive muons one at a time in the sample. The local magnetic field at the muon stopping site, i.e. the vector sum of the internal field intrinsic to the sample and any externally applied field, results in Larmor precession of the muon spin. The muon decays into a positron and two neutrinos after a mean lifetime of 2.2~$\mu$s, with the positron being emitted preferentially in the direction of the muon spin at the moment of decay. The positron events are recorded by detectors near the sample position. The main experimental quantity of interest is the asymmetry, defined as the normalized difference in positron events between a pair of detectors placed on opposite sides of the sample. Physically, the asymmetry is proportional to the projection of spin polarization of the muon ensemble along the axis defined by the detector pair. Thus, the internal magnetic field distribution in the sample can be probed via analysis of the time-dependent \muSR\ asymmetry~\cite{yaoua;b;msr11}. A helium gas flow cryostat was used to control the temperature between 2~K and 300~K. Asymmetry spectra were collected to a maximum time window of 9.7~$\mu$s. Fits to the \muSR\ spectra and other analyses were done with MUSRFIT~\cite{suter;physproc12} and home-built python software. The \muSR\ parameter $\alpha$ was calibrated using the spectra showing the fastest relaxation, allowing us to measure the true zero of the asymmetry spectrum directly. The usual calibration method using a weak transverse field in the paramagnetic state was not possible due to the high magnetic ordering temperature of \feo\ ($\sim$858~K).
	
\section{Results}	
\subsection{5 nm nanoparticles: TEM and Magnetometry}
\begin{figure}
	\includegraphics[width=70mm]{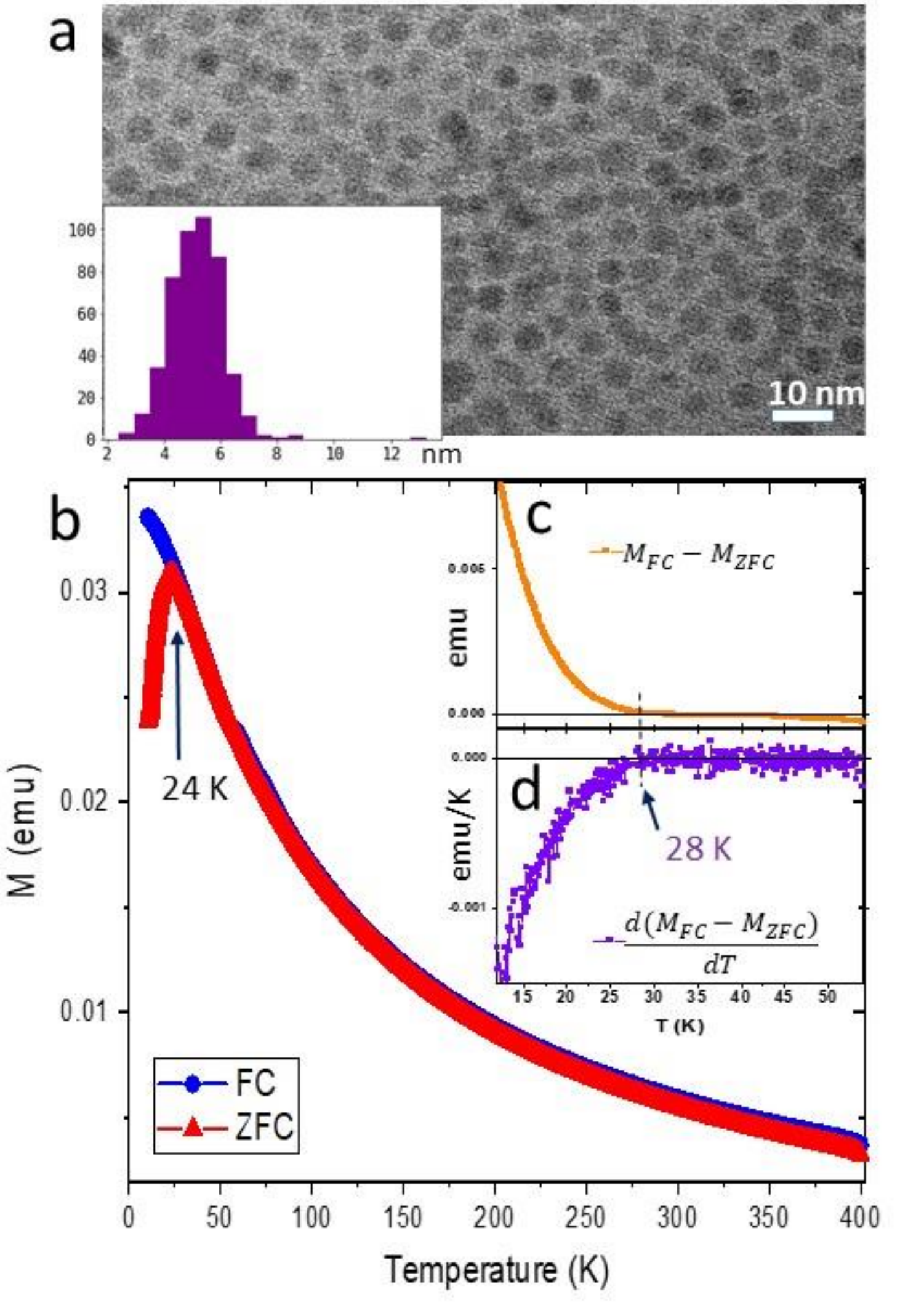}
	\caption{\label{fig:5nmTEM-VSM} (a) TEM image of a self assembly of 5~nm NPs deposited on a thin carbon membrane. Inset: Histogram of particle sizes yielding an average size of 5.16 $\pm$ 0.96~nm. (b) FC/ZFC curves collected under a field of 100~Oe (10 mT) over the temperature range 10 -- 400~K. (c) Magnetization difference $\Delta M=M_{\mathrm{FC}}-M_{\mathrm{ZFC}}$ and (d) its derivative with respect to temperature, $\mathrm{d}\Delta M/\mathrm{d}T$, plotted over the temperature range 10 -- 50~K.}
	
\end{figure}

We first present the magnetometry and TEM characterization of the 5~nm sample. After deposition onto a thin membrane, the NPs tend to self-assemble and form a close-packed hexagonal lattice, as indicated by the TEM image in Fig.~\ref{fig:5nmTEM-VSM}(a).
When the initial NP concentration is optimal, one can achieve a uniform monolayer of NPs, such as the one seen in Fig.~\ref{fig:5nmTEM-VSM}(a). A fine analysis of the particle size carried out on multiple TEM images containing about 465 NPs yields an average particle diameter of 5.16 $\pm$ 0.96~nm. Each particle is surrounded by an oleic acid ligand shell of approximately 1~nm in thickness, so that the average inter-particle distance is in fact closer to 6~nm or larger, as x-ray magnetic scattering measurements on identically prepared samples have revealed~\cite{chesn;mag18}. The formation of a vast, uniform, close-packed lattice is enabled by the narrow size distribution of the sample.

The FC/ZFC measurements shown in Fig.~\ref{fig:5nmTEM-VSM}(b) suggest a superparamagnetic (SPM) behavior where the associated blocking transition is indicated by the presence of a peak in the ZFC curve accompanied by the merging of the ZFC and FC curves. The narrowness of the peak exhibited by the ZFC curve reflects the narrowness of the particle size distribution. This result is consistent with FC/ZFC measurements on previous batches of 5~nm NPs prepared using the same method~\cite{chesn;jpconfs14,klomp;ieeem20}. The proper way to estimate the blocking temperature \Tb\ from magnetometry data has been subject to debate in recent years~\cite{bruve;jap15,lives;sr18}. We provide here two estimates of \Tb\ using the following two different criteria: 1) the location of the peak in the ZFC curve, as illustrated in Fig.~\ref{fig:5nmTEM-VSM}(b); and 2) the merging point of the FC/ZFC curves, where the difference $\Delta M(T)=M_{\mathrm{FC}}-M_{\mathrm{ZFC}}$ and its derivatives, $\mathrm{d}\Delta M/\mathrm{d}T$, $\mathrm{d}^2\Delta M/\mathrm{d}T^2$, etc., reach zero~\cite{lives;sr18}. As illustrated in Fig.~\ref{fig:5nmTEM-VSM}(c), $\Delta M(T)$ is the largest at the lowest temperature and decreases as $T$ increases until it reaches a merging point where $\Delta M=0$, above which it plateaus at zero (marking the SPM phase). Likewise, the magnitude of the derivative $\vert \mathrm{d}\Delta M/\mathrm{d}T \vert$ decreases when $T$ increases until plateauing at $\mathrm{d}\Delta M/\mathrm{d}T=0$, as seen in Fig.~\ref{fig:5nmTEM-VSM}(d). For the 5~nm NPs, the estimated blocking temperature using the ZFC peak is $T_B \approx 24$~K, whereas the merging criterion yields an estimated value of $T_B \approx 28$~K, somewhat higher but not significantly different.

The VSM data and the estimate of \Tb\ may be used to obtain further information about the dynamics of the nanospins in the SPM state. Based on the N\'eel-Arrhenius law~\cite{neel;adg49}, the mean time between two nanoparticle spin flips, called the N\'eel relaxation time $\tau_N$, is given by
\begin{align}\label{eq:tauNeel}
\tau_N=\tau_0\exp\left(\frac{E_A}{k_B T} \right),
\end{align}
where $\tau_0$ is the intrinsic nanoparticle spin fluctuation time (i.e., the magnetic reversal attempt time, which is also the value of $\tau_N$ at infinite temperature), $E_A$ is the activation energy or energy barrier for a nanospin flip, $k_B$ is the Boltzmann constant, and $T$ is the temperature of the material. For measurements where the sampling time $\tau_m$ is kept constant, the blocking temperature \Tb, delineating the SPM state from the blocked state, is the temperature for which $\tau_m = \tau_N$ such that
\begin{align}\label{eq:tau_m}
	\tau_m=\tau_0\exp\left(\frac{E_A}{k_B T_B} \right).
\end{align}
If $\tau_m$, \Tb, and $E_A$ are measured experimentally, an estimate of $\tau_0$ may then be obtained.

For our VSM measurements, $\tau_m \approx 1$~s, and the blocking temperature for the 5~nm NPs is $T_B \approx 24$~K. As explained below, the \muSR\ data provides an estimate of the activation energy of the 5~nm NPs to be $E_A \approx 1.4 \times 10^{-21}$~kJ. From these values, we obtain the ratio $E_A/(k_B T_B )\approx4.2$ and $\tau_0\approx1.4\times10^{-2}$~s (14~ms). However, this estimated value for $\tau_0$ is several orders of magnitude higher than values found in the literature, which typically range from $10^{-9} - 10^{-10}$~s~\cite{majet;jpd06} or even as short as $10^{-13}$~s~\cite{perig;aprev15}. This large discrepancy may be explained by missing terms in the expression for the energy barrier, such as magnetic interaction terms as described by W.F. Brown~\cite{brown;pr63}. 

Here we attempt a simple correction to the energy barrier  magnetic interaction between the nanospins and the applied magnetic field (Zeeman effect): $E_M= \Delta \vec{M}\cdot \vec{H}$  where $\Delta \vec{M}$ is the change in magnetic moment carried by each NP caused by a flip and $\vec{H}$ is the applied magnetic field. Assuming a full up-down flip occurs along the direction of the applied field, $E_M= 2M_s H$ where $M_s$ is the magnetization at saturation for an individual NP. The adjusted energy barrier is, to a first approximation, $\Delta E= E_A+ E_M$ and the adjusted fluctuation time $\tau_0^*$ is then given by
\begin{align}\label{eq:taustar}
	\tau_0^*=\tau_m\exp\left(-\frac{E_A+E_M}{k_B T_B} \right).
\end{align}

The magnitude of $M_s$ may be estimated by using the volume of each NP and the density of magnetic ions in the material. Based on previous crystallographic studies of these samples~\cite{klomp;ieeem20}, the cubic lattice parameter is 8.35~\AA\ and the atomic density is about 13.7 \feo\ units per nm$^3$. Accounting for combined spins carried by the Fe$^{2+}$ and Fe$^{3+}$ ions in the spinel structure~\cite{cai;jap14}, the saturated magnetization density is about 55~$\mu_B$/nm$^3$. Assuming a spherical NP shape with a radius of 2.6~nm, we have $M_s=4000$~$\mu_B$. The applied field during the VSM measurement was $H=100$~Oe (10~mT). The resulting value for the Zeeman magnetic interaction then comes to $E_M=7.4\times 10^{-22}$~J. The adjusted energy ratio $\Delta E / k_B T_B \approx 6.46$ leads to an adjusted attempt time $\tau_0^* \approx 1.5 \times 10^{-3}$ (1.5~ms), only a factor of 10 smaller than the initial estimate of $\tau_0$.

Evidently, this simple model does not appear to reflect accurately the actual energy barrier $\Delta E$ experienced by the NPs during the VSM measurements, but may be refined by including additional energy terms, such as the ferromagnetic exchange energy, the magnetostatic energy, the crystalline and shape anisotropy energy, and interparticle magnetic interactions. The contributions from these factors could be derived from various models, such as the Stoner-Wohlfarth model, the Gilbert equations, and the Fokker-Planck equations, as suggested theoretically by Brown~\cite{brown;pr63} and experimentally by more recent works~\cite{majet;jpd06,torre;jap15}. In the scope of our present paper, instead of attempting to refine our estimate of the energy barrier experienced by the NPs, we show in the following how \muSR\ provides a unique and reliable way to probe spin fluctuations in magnetic NP systems, making the technique highly complementary to bulk magnetometry.

\subsection{5 nm nanoparticles: \muSR}
\begin{figure}
	\includegraphics[width=70mm]{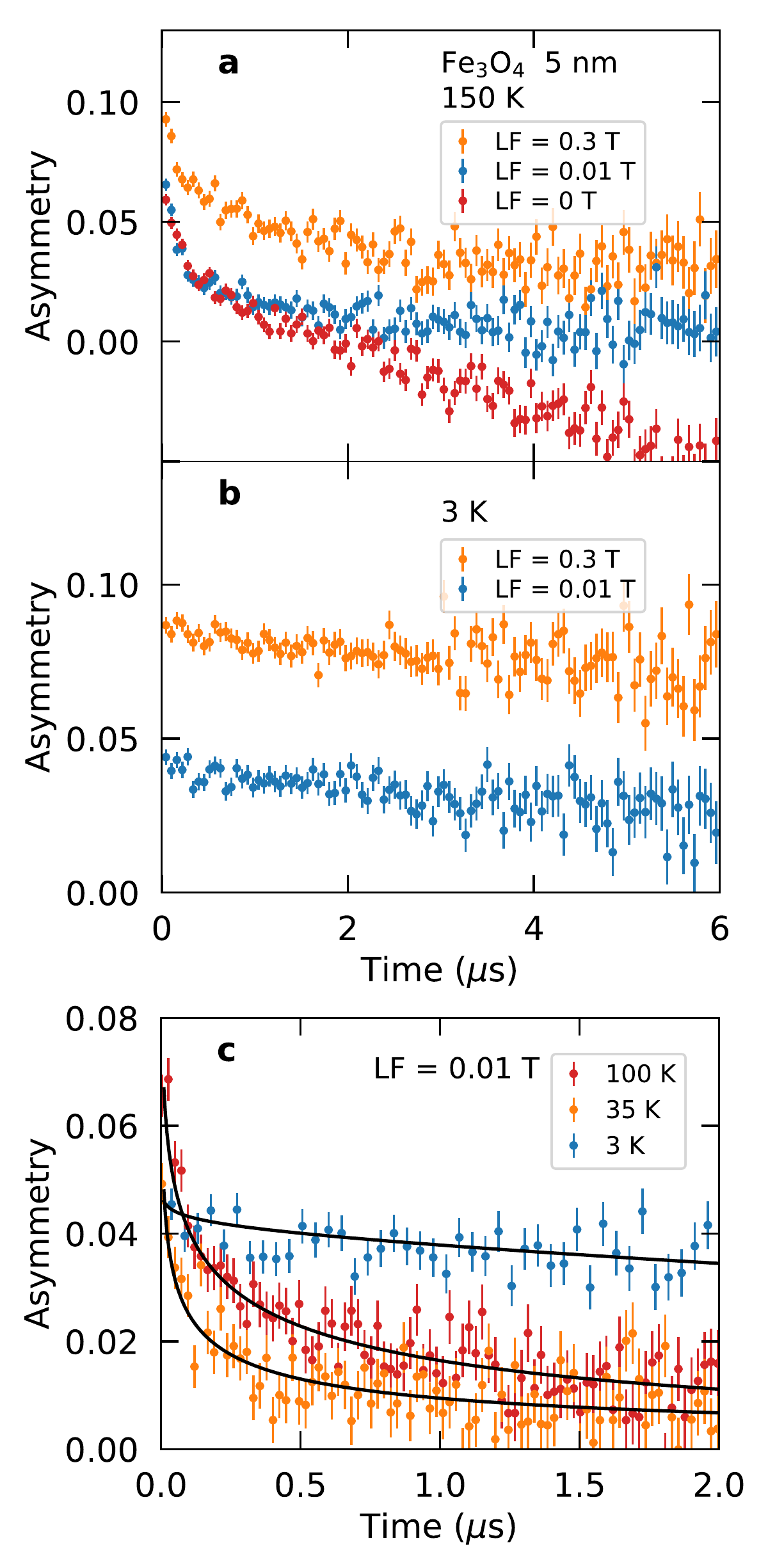}
	\caption{\label{fig:musr5nm} (a) \muSR\ asymmetry spectra for 5 nm \feo\ NPs collected at 150~K in various applied longitudinal fields (LF). (b) Same as (a) but for data collected at 3~K. (c) Asymmetry spectra collected in a weak longitudinal field of 0.01~T at 100~K (superparamagnetic), 35~K (mixed superparamagnetic and blocked states), and 3~K (blocked). The solid black curves are fits described in the main text.}
		
\end{figure}

To obtain additional insight into the behavior of this system, we now turn to the \muSR\ characterization of the 5~nm NPs. Fig.~\ref{fig:musr5nm} displays \muSR\ asymmetry spectra collected at representative temperatures and field conditions. 
In panel (a), we display spectra collected at 150~K in various values of an externally applied longitudinal field (LF), i.e. an applied field directed along the initial muon spin polarization direction. Significant relaxation of the asymmetry is observed even in the strongest applied field LF = 0.3~T, indicating that the intrinsic magnetic field at the muon site is dynamically fluctuating. This is expected, since the sample is in the SPM state at this temperature. We note that the spectrum collected with zero applied field appears to consist of two asymmetry components, one that relaxes rapidly within the first $\sim$2~$\mu$s, and the other which displays a more gentle relaxation. The slowly relaxing asymmetry from the latter component is fully recovered when a modest field of LF = 0.01~T is applied, as seen by the blue symbols in Fig.~\ref{fig:musr5nm}(a), indicating that the muons contributing to this asymmetry component experience a very weak internal field. We attribute this slowly relaxing asymmetry component to muons stopping in the ligand shells around the NPs, where the intrinsic field is much weaker. The faster relaxing asymmetry component arises from muons stopping within the NPs, where the large intrinsic field dominates the externally applied field. Since the asymmetry component corresponding to muons stopping within the NPs is of primary interest here, we calibrated the zero-asymmetry level (i.e. the $\alpha$ parameter) using the rapidly relaxing spectra in LF = 0.01~T. Hence, the negative asymmetry observed for LF = 0 in Fig.~\ref{fig:musr5nm}(a) is artificial.

The persistent relaxation of the spectra shown in Fig.~\ref{fig:musr5nm}(a) demonstrates the presence of SPM dynamics at 150~K, as expected. Similar measurements performed at 3~K show strongly contrasting behavior, as seen in Fig.~\ref{fig:musr5nm}(b). In this case, the spectra for LF = 0.01~T and 0.3~T both show very little relaxation with time. Furthermore, the overall asymmetry level for LF = 0.3~T is much higher than for 0.01~T. This ``decoupling'' of the asymmetry with increasing LF, together with the lack of strong relaxation, demonstrate that the local internal field for the vast majority of the muons is static, consistent with the blocked state expected for this behavior. Interestingly, the slight relaxation still observed at 3~K as seen in Fig.~\ref{fig:musr5nm}(b) indicates that a very small fraction of the muons experience dynamically fluctuating fields even at 3~K, suggesting that some small portion of the sample remains in the SPM state at this low temperature. Nevertheless, the data shown in Fig.~\ref{fig:musr5nm}(a) and (b) clearly reveal that the SPM behavior at high temperature gives way to a blocked state at low temperature, fully in accord with the expectations for this sample.

To gain a better understanding of the blocking transition, we collected \muSR\ data at a series of temperatures between 3~K and 150~K with LF = 0.01~T. The application of this weak field allows us to decouple the slowly relaxing asymmetry arising from muons stopping in the ligand shell so that any remaining relaxation can be related directly to the behavior of the NPs. Fig.~\ref{fig:musr5nm}(c) displays the spectra at the representative temperatures 100~K, 35~K, and 3~K. At 100~K, the asymmetry relaxes steadily and approaches zero due to the SPM fluctuations. On the other hand, the spectrum collected at 3~K [the same one as shown in Fig.~\ref{fig:musr5nm}(b)], remains relatively constant in time, as expected for NPs in the blocked state. It is also noteworthy that at $t=0$, the 100~K spectrum has a higher value of the initial observed asymmetry than does the 3~K spectrum. This is due to the well-known phenomenon by which the \muSR\ spectra for samples with randomly oriented, static magnetic domains (individual NPs in this case) contain two components: a ``two-thirds'' asymmetry component with oscillations and/or strong damping due to internal field components that are perpendicular to the initial muon spin direction, and a ``one-third'' tail showing greatly reduced relaxation arising from the internal field components that are parallel to the muon spin and therefore cause no Larmor precession. In the present case, only the one-third tail is visible for the spectrum collected at 3~K; the two-thirds component apparently cannot be resolved within the time resolution of the detectors due to very large internal fields and/or a broad field distribution at the muon site. At 100~K, the NPs undergo SPM fluctuations that do not result in a loss of the two-thirds component, explaining the higher initial ($t=0$) asymmetry value for 100~K.

The spectrum collected at 35~K displays more rapid relaxation than either of the other two spectra displayed in Fig.~\ref{fig:musr5nm}(c). Considering that this is close to \Tb\ = 24~K observed in the magnetometry data, this enhanced relaxation rate can be attributed to slow dynamics of the SPM fluctuations near the blocking transition, very similar to the enhanced \muSR\ relaxation due to critical dynamics observed in the vicinity of any conventional magnetic ordering transition in bulk materials~\cite{frand;prm20,yaoua;b;msr11,uemur;ms99}. In addition, close inspection of the 35~K spectrum reveals that the initial observable asymmetry lies between those of the 100~K and 3~K spectra, suggesting that the sample is in an intermediate state at 35~K, with part of the sample blocked and the rest in the SPM state. 

\begin{figure}
	\includegraphics[width=70mm]{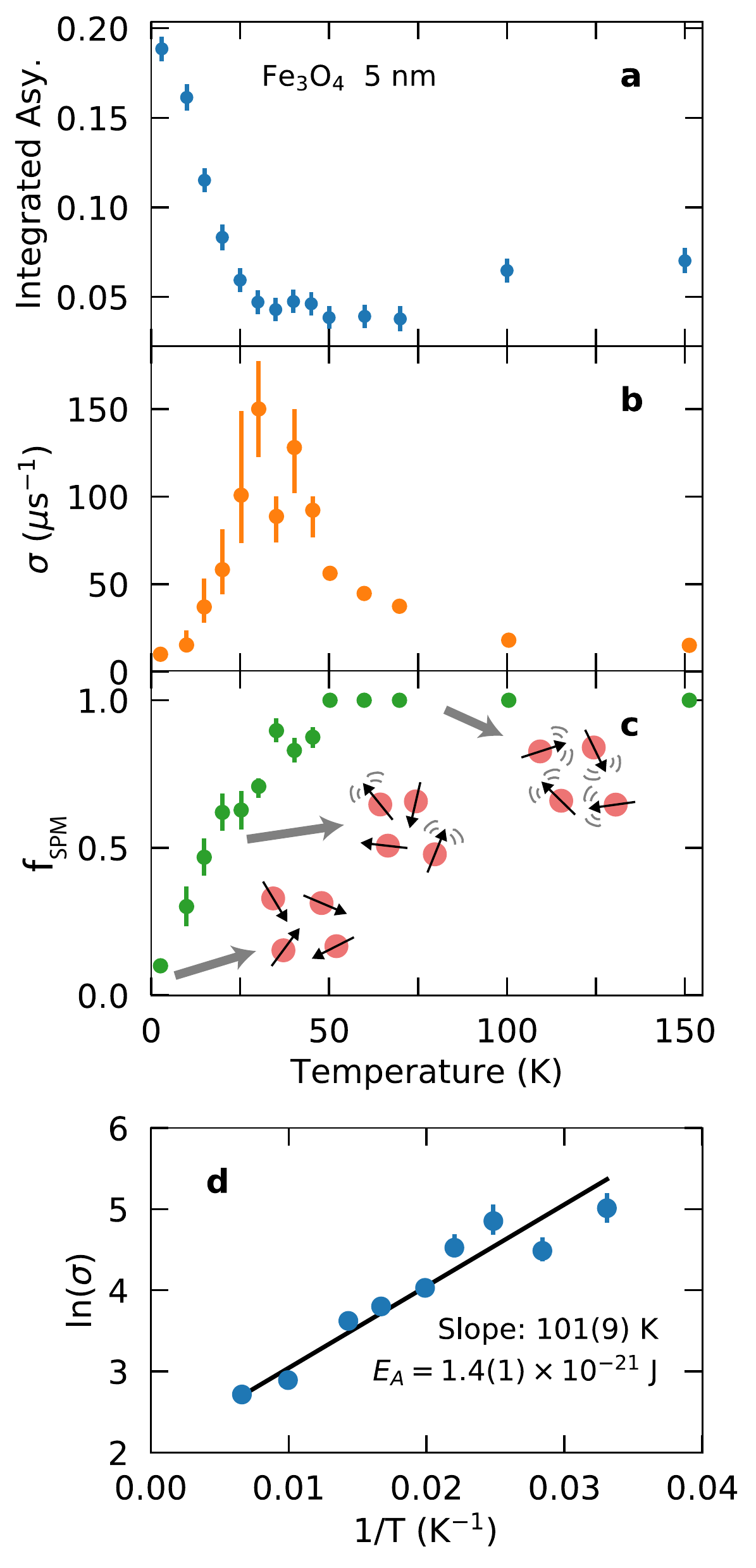}
	\caption{\label{fig:5nmanalysis} (a) Integrated \muSR\ asymmetry as a function of temperature for the 5 nm NPs. (b) Temperature dependence of the superparamagnetic relaxation rate $\sigma$ as determined by least-squares fits. (c) Volume fraction of the sample exhibiting superparamagnetic behavior as a function of temperature, obtained from the least-squares fits. The schematic diagrams illustrate (from right to left) the fully superparamagnetic state, mixed state, and fully blocked state. (d) Logarithm of the superparamagnetic relaxation rate $\sigma$ plotted against $1/T$. The solid line represents a fit using the activation model described in the text.}
\end{figure}

We now proceed with a more quantitative analysis of the \muSR\ data to probe the blocking transition. First, we display in Fig.~\ref{fig:5nmanalysis}(a) the results of integrating the experimental asymmetry spectra over the first 8~$\mu$s for each temperature point at which data were collected. 
The most striking feature is a sharp rise in the integrated asymmetry beginning below 30~K, which results from an increasing portion of the sample entering the blocked state and consequently experiencing greatly reduced relaxation. This simple quantification of the \muSR\ asymmetry spectra is an effective way to gain a rough estimate of \Tb.

More detailed understanding can be gained by fitting a mathematical model to the asymmetry spectra using least-squares optimization. Since individual muons act as local probes of their immediate surroundings, regions of the sample in distinct magnetic states (such as SPM and blocked states) will make distinct contributions to the total observed asymmetry. Accordingly, we use the following two-component model to describe the asymmetry spectra collected in LF = 0.01~T: 
\begin{align}\label{eq:fitFunc}
	a(t) = a_{0}\left[ f_{\mathrm{SPM}}\times e^{-(\sigma t)^{\beta_1}}+\frac{1-f_{\mathrm{SPM}}}{3}\times e^{-(\lambda t)^{\beta_2}}\right],
\end{align}
where $a_0$ is the total initial asymmetry observed in the SPM state, \fSPM\ is the volume fraction of the sample in the SPM state, $\sigma$ is the SPM relaxation rate, $\lambda=1/T_1$ is the conventional electronic relaxation rate observed in the blocked state due to intra-particle spin flip processes, and $\beta_1$ and $\beta_2$ are exponential ``stretching'' parameters for the SPM and blocked states, respectively. The 3 in the denominator of the coefficient on the second term reflects the fact that we only observe the one-third tail of the asymmetry from the blocked state. This model was successfully used in a \muSR\ study of $\gamma$-Fe$_2$O$_3$ nanoparticles~\cite{rebbo;prb07} and is justified by an earlier theoretical work~\cite{lord;jpconfs05}. For ideally monodisperse nanoparticles with a single fluctuation rate, $\beta_1=0.5$ and $\beta_2=1$. If these assumptions are not met, $\beta_1$ and $\beta_2$ will typically deviate to become lower than these ideal values~\cite{lord;jpconfs05}.

We performed least-squares fits to the temperature-dependent \muSR\ spectra using this model with $a_0$, $\beta_1$, and $\beta_2$ set as global parameters common for all spectra, while \fSPM, $\sigma$, and $\lambda$ were allowed to vary with temperature. The resulting fits provided excellent matches to the data, as seen by the black curves in Fig.~\ref{fig:musr5nm}(c). The value of $a_0$ came to 0.13, which is less than the full asymmetry of $\sim$0.2 - 0.25 expected at TRIUMF due to the decoupled component from muons landing in the ligand shells described previously. $\beta_1$ refined to 0.25, which is consistent with the finite size distribution of the sample. $\beta_2$ tended to remain close to the ideal value of 1, so we fixed it to 1 for simplicity. The temperature-dependent values of $\lambda$ all remained fairly small ($\lesssim 0.3~\mu$s).

The most important parameters for understanding the blocking transition in the sample are $\sigma$ and \fSPM, which are displayed versus temperature in Fig.~\ref{fig:5nmanalysis}(b) and (c). The SPM relaxation rate $\sigma$ in panel (b) displays a dramatic peak centered around 30~K indicative of the blocking transition, consistent with the change of slope in the integrated asymmetry seen in Fig.~\ref{fig:5nmanalysis}(a) and the divergence of the FC and ZFC magnetization curves shown in Fig.~\ref{fig:5nmTEM-VSM}(b). Inspecting \fSPM\ in panel (c), we see that the SPM fraction is below 0.1 at the lowest measured temperature of 3~K, indicating that at least 90\% of the sample is fully blocked at this temperature. With increasing temperature, \fSPM\ rises steadily, passing the midpoint (i.e. \fSPM~=~0.5) around 17~K and finally reaching 1 for 50~K and above, indicating that the full sample volume is SPM at these temperatures. Thus, for the entire temperature region between approximately 3~K and 50~K, the data suggest sample is in a mixed state. In this mixed state, some regions of the sample exhibit SPM behavior while other regions are in the blocked state, as illustrated by the schematic diagrams in Fig.~\ref{fig:5nmanalysis}(c). Based on these results, we conclude that the blocking transition occurs gradually over a finite temperature window between approximately 3~K and 50~K. From this, we can define three convenient temperatures related to the blocking transition: the onset blocking temperature, which is the highest temperature at which a nonzero volume of the sample is in the blocked state; the midpoint blocking temperature, at which 50\% of the sample volume is in the blocked state; and a completion blocking temperature, at which 100\% of the sample is in the blocked state. For the current sample, the onset, midpoint, and completion blocking temperatures are $\gtrsim 45$~K, $\sim 16.6$~K, and $\lesssim 3$~K, respectively. We stress that the nature of \muSR\ as a local probe imbues this technique with a rather unique ability to quantify the segregation or coexistence of distinct magnetic states as seen here, providing complementary information to bulk probes such as magnetometry.  

We can obtain microscopic details about the SPM properties of the sample through further analysis of the relaxation rate $\sigma$. The monotonic decrease of $\sigma$ as the temperature is raised from 30~K (roughly where the peak in $\sigma$ occurs) to 150~K results from thermal activation of nanoparticle spin flip processes and can be modeled by~\cite{rebbo;prb07} 
\begin{align}\label{eq:Arrhenius}
	\sigma(T)=\sigma_0 \exp\left({E_A/k_{\mathrm{B}}T}\right),
\end{align}
where $T$ is the temperature in kelvin, $\sigma_0$ is a type of intrinsic relaxation rate related to the asymptotic dynamics at high temperature, $E_A$ is the activation energy for flipping the net magnetic moment of an individual nanoparticle, and $k_{\mathrm{B}}$ is the Boltzmann constant. This is most conveniently displayed as a logarithmic plot of $\sigma$ versus inverse temperature, as shown in Fig.~\ref{fig:5nmanalysis}(d). Eq.~\ref{eq:Arrhenius} appears as a line in this type of plot, with the slope equal to $E_A/k_{\mathrm{B}}$. A fit to the data is shown by the black line in Fig.~\ref{fig:5nmanalysis}(d). The best-fit value of $E_A$ is $1.4(1)\times 10^{-21}$~J, corresponding to a magnetic anisotropy constant $K=E_A/V$ (with $V$ the NP volume) of $19 \pm 11$~kJ/m$^3$. While experimental limitations and the finite size distribution of our sample preclude a highly precise calculation of $K$ from our data, our estimate is nevertheless generally consistent with other values reported in the literature~\cite{rebbo;prb07,tacke;jmmm08,perig;aprev15,dobos;jmmm17}, lending confidence to this result.

The best-fit value of $\sigma_0$ can further be used to estimate the previously mentioned magnetic reversal attempt time $\tau_0$ in the limit of high temperature using the relationship~\cite{lord;jpconfs05,rebbo;prb07}
\begin{align}\label{eq:sigma0}
	\sigma_0 = 4(\gamma_{\mu}\Delta B)^2 \tau_0,
\end{align}
where $\gamma_{\mu} = 851.6$~$\mu$s$^{-1}$T$^{-1}$ is the muon's gyromagnetic ratio and $\Delta B$ is the half width at half maximum of the typical distribution of internal fields at the muon stopping sites. Since the instantaneous internal field is due primarily to the ferrimagnetic order of magnetite within each nanoparticle, we can estimate $\Delta B$ from the transverse damping rate of the spontaneous asymmetry oscillations observed in bulk \feo\ (data not shown)~\cite{dunsi;prb96,frand;prm20}. This yields an estimate of $\tau_0$ of $2.8(5)\times 10^{-10}$~s, likewise in line with published estimates for magnetite and related nanoparticle systems
~\cite{morup;prl94,tronc;jmmm00,rebbo;prb07,tacke;jmmm08,perig;aprev15}.

As a final check for self-consistency, we insert the experimentally determined values of $E_A$ and $\tau_0$, together with $\tau_m\approx 10^{-6}$~s for \muSR, into Eq.~\ref{eq:tau_m} to calculate \Tb. The result is $T_B \sim 12$~K, which falls comfortably within the range of temperatures over which the sample is transitioning between the SPM and blocked states. This consistency supports the reliability of the estimates of the microscopic parameters obtained from the \muSR\ analysis. It also highlights the value of using \muSR\ to estimate $\tau_0$ directly from the temperature-dependent relaxation rate, providing an independent calculation in addition to the more frequently used relationship in Eq.~\ref{eq:tau_m}.

\subsection{20 nm nanoparticles: TEM and Magnetometry}
\begin{figure}
	\includegraphics[width=70mm]{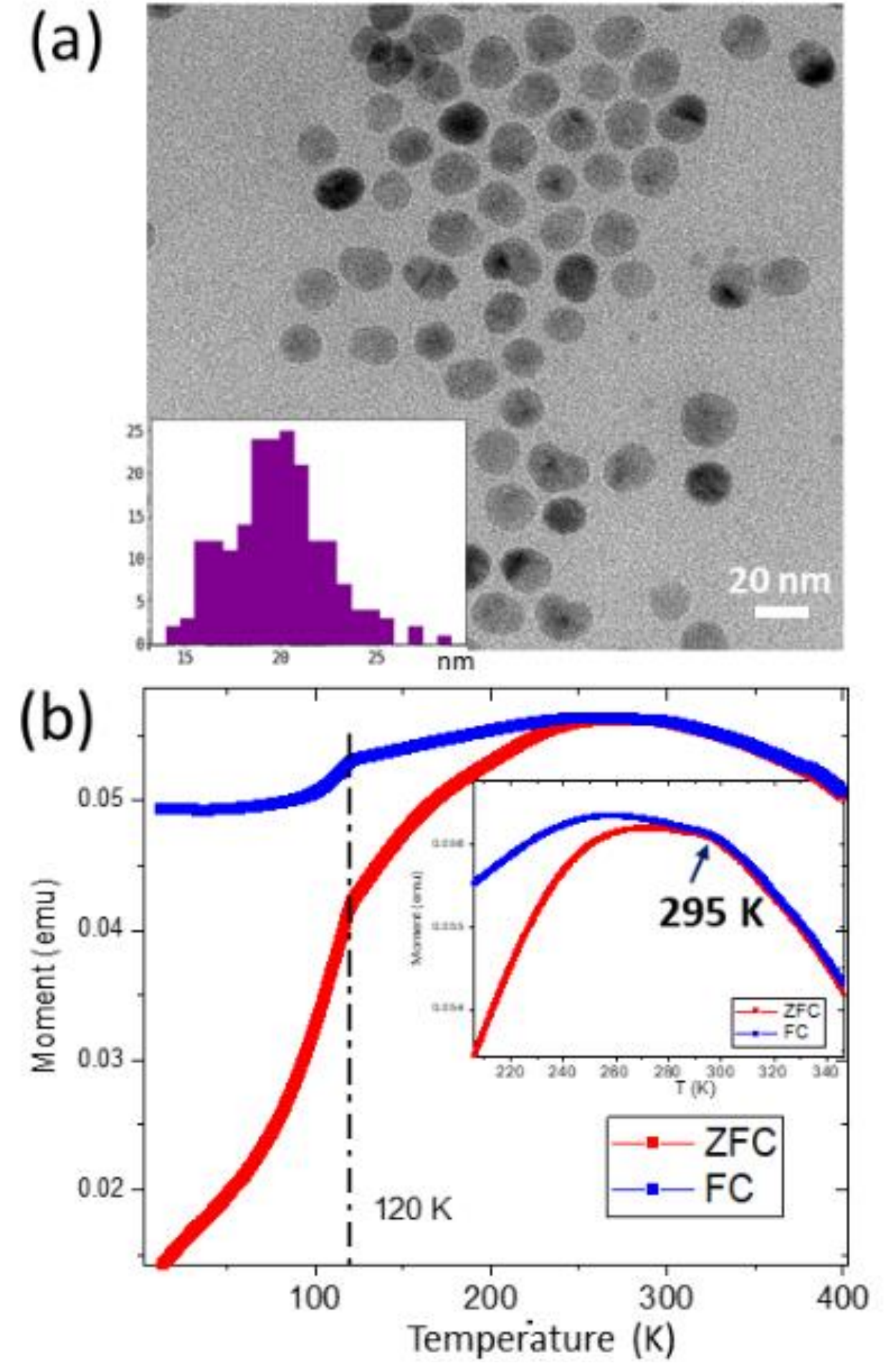}
	\caption{\label{fig:20nmTEM-VSM} (a) TEM image of a 20 nm NP self-assembly deposited on a thin carbon membrane. Inset: Histogram of particle sizes yielding an average size of $19.22 \pm 2.37$~nm. (b)  FC/ZFC curves collected under a field of 100 Oe (10 mT). Inset: Zoomed-in view of the blocking transition region.
	}		
\end{figure}

Equivalent characterizations using TEM, magnetometry, and \muSR\ were carried out on the sample of 20 nm \feo\ NPs. When deposited on a thin membrane for TEM imaging, the 20 nm NPs tend to self-assemble in a hexagonal lattice, similar to the 5 nm NPs. The TEM image in Fig.~\ref{fig:20nmTEM-VSM}(a) indicates the formation of monolayers of closely packed 20 nm NPs distributed in sparse islands throughout the film. 
With optimal NP concentration, one may achieve a uniform monolayer coverage, with a few particles stacking on top of each other. A fine analysis of the particle size carried out on multiple TEM images including about 265 NPs yields an average particle size of 19.23 $\pm$ 2.37~nm. 

The FC/ZFC curves in Fig.~\ref{fig:20nmTEM-VSM}(b) indicate that the 20~nm NPs also exhibit superparamagnetism, with a blocking transition occurring at $T_B \approx 295$~K using the ZFC peak criterion and $T_B \approx 290$~K using the merging point criterion, again not significantly different. It has been suggested~\cite{li;sr17} that 20 nm may be the upper NP size limit for SPM to occur, but in any case, the commercially supplied 20 nm NPs studied here unambiguously display SPM above room temperature. Another remarkable feature observed in both the FC and ZFC curves is a kink at about 120~K, which we interpret as a signature of the Verwey transition, a magneto-crystalline transition normally occurring in bulk \feo\ at $T_{\mathrm{V}}=125$~K~\cite{verwe;phy41}. This transition has been previously observed in FC/ZFC curves collected on larger \feo\ NPs. One study on 50-100 nm NPs shows the occurrence of a clear peak at 125 K~\cite{klomp;ieeem20}. Another study shows a kink in the FC/ZFC curves located at lower temperatures, 16~K for 50 nm NPs and 98~K for 150 nm NPs~\cite{goya;jap03}. For NPs smaller than 20 nm, FC/ZFC studies have established the complete vanishing of the Verwey transition~\cite{klomp;ieeem20}. It is therefore interesting to observe a sign of the Verwey transition in the present 20 nm NP material.

Following the earlier analysis of the VSM data to obtain an estimate of the attempt time $\tau_0$, we calculate the ratio $E_A / k_B T_B$ to be $\sim$5.7, using $T_B=290$~K and $E_A=2.3 \times 10^{-20}$~J as obtained from the \muSR\ data (see below). This yields an attempt time of $\tau_0 \approx 3.2 \times 10^{-3}$~s (3.2~ms), once again far too large a value to be realistic. Using the energy correction method described previously, we find that the saturated magnetization for the 20~nm NPs is $M_s\approx 204000$~$\mu_B$, leading to a Zeeman energy of $E_M \approx 3.8\times 10^{-20}$J. The adjusted energy ratio $\Delta E / k_B T_B \approx 15.2$ corresponds to an adjusted attempt time $\tau_0^*\approx 2.5 \times 10^{-7}$~s (250 ns). This is about 4 orders of magnitude smaller than the initial estimate of $\tau_0$, but still too large with respect to the expected value, meaning not all of the relevant interactions have been properly modeled.

\subsection{20 nm nanoparticles: \muSR}
\begin{figure}
	\includegraphics[width=70mm]{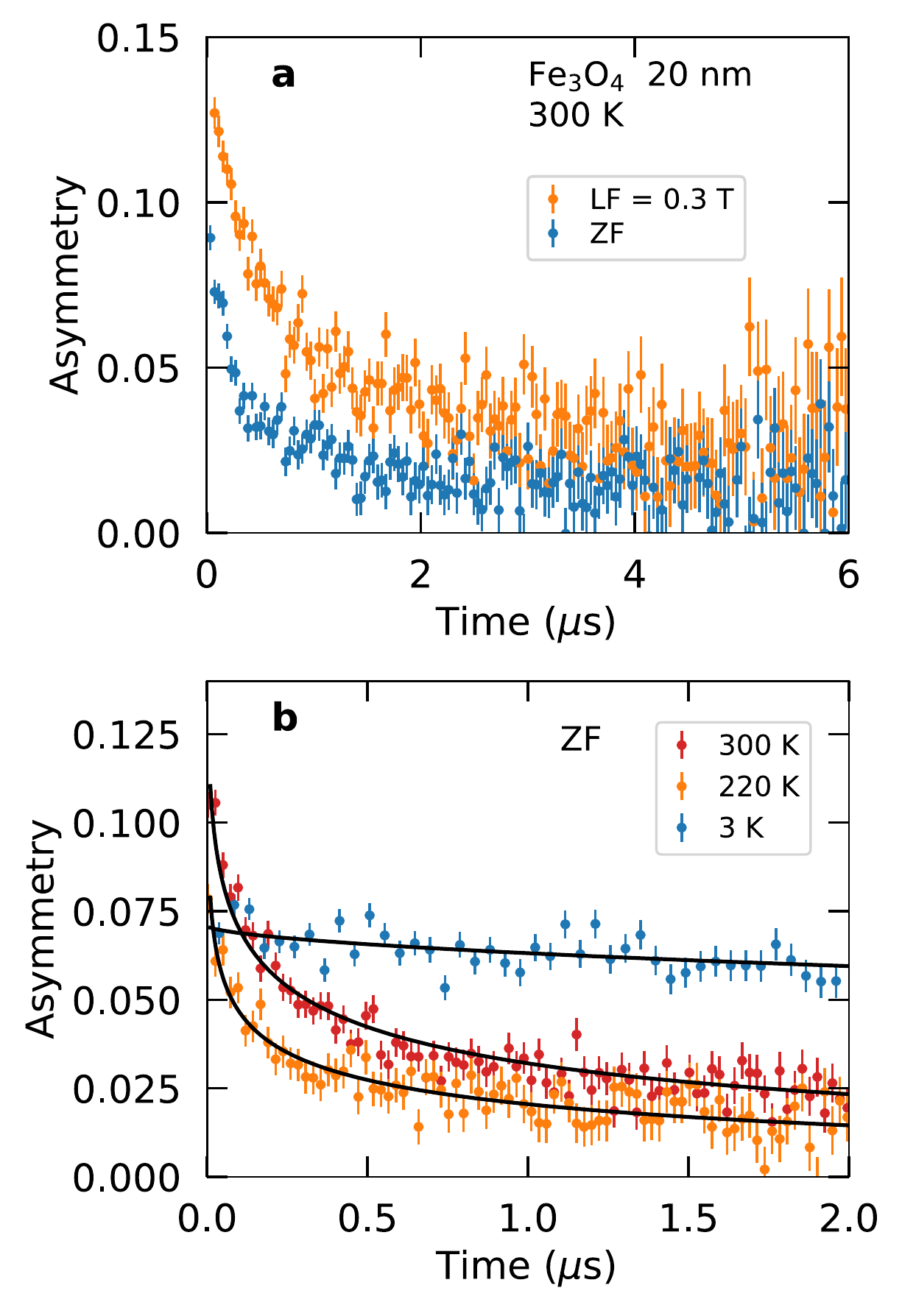}
	\caption{\label{fig:musr20nm} (a) \muSR\ asymmetry spectra for 20 nm \feo\ NPs collected at 300~K in zero field (ZF) and LF = 0.3~T. (b) ZF asymmetry spectra at 300~K (mostly superparamagnetic), 220~K (mixed superparamagnetic and blocked states), and 3~K (blocked). The solid black curves are fits described in the main text.
	}		
\end{figure}

The \muSR\ data collected on the 20 nm NPs are qualitatively similar to those of the 5 nm NPs, confirming a broad transition from superparamagnetism at high temperature to the blocked state at low temperature. Spectra collected at 300~K in zero field (ZF) and LF = 0.3~T are shown in Fig.~\ref{fig:musr20nm}(a), where the strong relaxation in the LF spectrum confirms the prevalence of SPM dynamics.
Interestingly, the ZF spectrum does not show the slowly relaxing component observed in the 5 nm data. This is likely because the much greater volume of the NPs in this sample compared to the 5 nm sample results in a much larger fraction of muons landing inside the NPs rather than in the ligand shell, so the weak relaxation from muons stopping in the ligand shell is negligible. ZF spectra at representative temperatures are shown in Fig.~\ref{fig:musr20nm}, where we once again observe fast relaxation at high temperature (300~K) in the SPM state, nearly negligible relaxation in the blocked state at low temperature (3~K), and intermediate behavior at 220~K suggesting a mixed state with both SPM and blocked regions of the sample.

\begin{figure}
	\includegraphics[width=70mm]{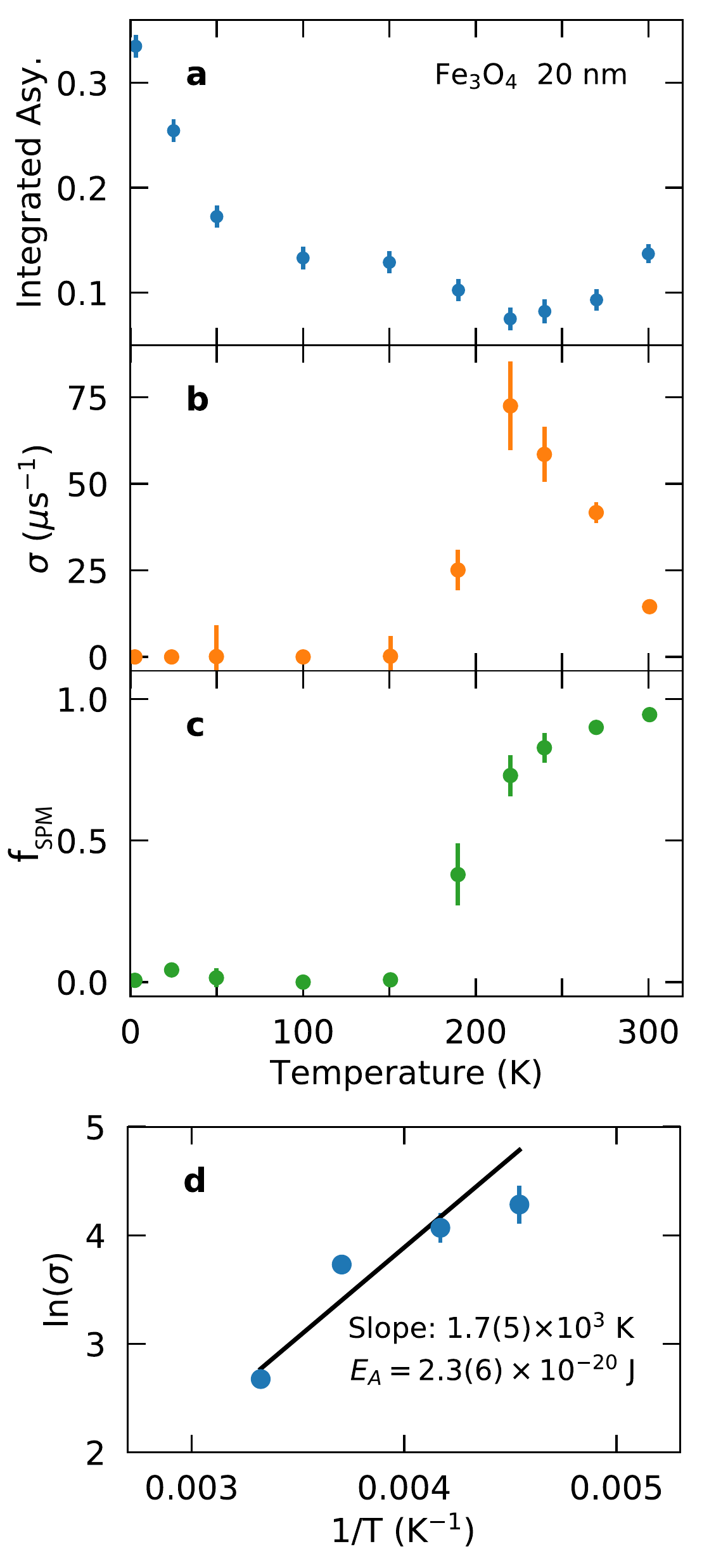}
	\caption{\label{fig:20nmanalysis} (a) Integrated \muSR\ asymmetry as a function of temperature for the 20 nm NPs. (b) Temperature dependence of the superparamagnetic relaxation rate $\sigma$ as determined by least-squares fits. (c) Volume fraction of the sample exhibiting superparamagnetic behavior as a function of temperature. (d) Logarithm of the superparamagnetic relaxation rate $\sigma$ plotted against $1/T$. The solid line represents a fit using the activation model described in the text.
	}		
\end{figure}

We carried out the same quantitative analysis for the 20 nm data as we did for the 5 nm data. The integrated asymmetry is plotted as a function of temperature in Fig.~\ref{fig:20nmanalysis}(a). 
A minimum is observed around 220~K, with the integrated asymmetry increasing monotonically as the temperature is lowered below 220~K. As before, this indicates that a significant fraction of the sample has entered the blocked state as the temperature is reduced beyond this point. 

Eq.~\ref{eq:fitFunc} was once again used to model the ZF spectra, with representative fits shown as the black curves in Fig.~\ref{fig:musr20nm}. The globally refined value of $a_0$ came to 0.21, $\beta_1$ refined to 0.25 as in the 5 nm case, and $\beta_2$ refined to 0.6. The temperature dependence of $\sigma$ and \fSPM\ are displayed in Fig.~\ref{fig:20nmanalysis}(b) and (c), where we again observe a prominent peak in $\sigma$ corresponding to the blocking transition and a gradual evolution of \fSPM, revealing a wide temperature range over which the mixed state with coexisting SPM and blocked regions exists in the sample. Interestingly, even at the highest temperature measured (300~K), the fits indicate that approximately 5-10\% of the sample is already in the blocked state. This fraction could plausibly be even higher, but without a reference measurement fully in the SPM state, there is some difficulty in unambiguously separating the two components in Eq.~\ref{eq:fitFunc} at high temperature. Regardless, the \muSR\ data indicate that the onset of the blocking transition in the \muSR\ time scale is above room temperature and therefore higher than the $T_B \approx 295$~K estimated from the magnetometry data in Fig.~\ref{fig:20nmTEM-VSM}. From the $f_{\mathrm{SPM}}$, the onset, midpoint, and completion blocking temperatures for the 20~nm sample are $> 300$~K, $\sim 200$~K, and $\sim 150$~K, respectively.

Fig.~\ref{fig:20nmanalysis}(d) shows the result of the thermal activation analysis carried out on the best-fit values of $\sigma$ using Eq.~\ref{eq:Arrhenius}. With only four data points comprising the high-temperature side of the peak in $\sigma$ [see Fig.~\ref{fig:20nmanalysis}(b)], our ability to determine the activation energy, anisotropy, and intrinsic nanoparticle spin fluctuation time reliably is quite limited compared to the 5 nm results. Pressing forward nonetheless, we estimate the activation energy to be $E_A = 2.3(6) \times 10^{-20}$~J, the anisotropy constant to be $K = 6.1 \pm 2.9$~kJ/m$^3$, and the intrinsic nanoparticle spin fluctuation time to be $\tau_0 \sim 10^{-11}$~s. We note that the estimated uncertainty for $\tau_0$ was larger than the calculated value itself, so this result should be considered cautiously. Nevertheless, inserting these values into Eq.~\ref{eq:tau_m} yields an estimated \Tb\ of 143~K, which is no more than a factor of $\sim$2 different from the observed blocking temperature.

\section{Discussion and Conclusion}	
The \muSR\ and magnetometry results presented here provide valuable information about the magnetic properties of \feo\ nanoparticles. Key quantities related to the magnetism in both samples are listed in Table~\ref{table:props}.
\begin{table}
	\ra{1.3}
	\caption{Selected magnetic properties of 5 and 20 nm \feo\ nanoparticles. The various estimates of the blocking temperature \Tb\ and the magnetic reversal attempt time $\tau_0$ are explained in the main text.} 
	\centering 
	\begin{tabular}{l c c} 
		& \multicolumn{2}{c}{Particle size} \\
		\cline{2-3} 
		& 5 nm & 20 nm \\  
		\hline
		\hline
		VSM results & & \\
		$T_B$ ZFC peak (K) & 24 & 295  \\
		$T_B$ merging (K) & 28 & 290  \\
		$E_M$ at 10 mT ($10^{-21}$ J) & 0.74 & 37  \\
		$\tau_0^{*}$ (s) & $1.5 \times 10^{-3}$ & $2.5 \times 10^{-7}$ \\
		\hline
		\muSR\ results & & \\
		$T_{B}$ onset (K)& 45 & $>$300 \\
		$T_{B}$ midpoint (K) & 16.6 & 200\\
		$T_{B}$ completion (K) & $<$3 & 150\\
		$E_{A}$ ($10^{-21}$ J) & 1.4 $\pm$ 0.1 & 23 $\pm$ 6 \\
		$K$ (kJ/m$^3$) & 19 $\pm$ 11 & 6.1 $\pm$ 2.9 \\
		$\tau_0$ (s) &$(0.28 \pm 0.05)\times 10^{-10}$ & $10^{-11\dagger}$ \\
		\hline
		\multicolumn{3}{l}{$^{\dagger}$Calculated uncertainty is larger than the value itself;}\\
		\multicolumn{3}{l}{ upper limit is $10^{-9}$~s.}
	\end{tabular}
	\label{table:props} 
\end{table}
Both techniques confirm the trend that with increasing particle size, the blocking temperature increases and the magnetic anisotropy decreases. This can be understood as a consequence of the smaller surface-to-volume ratio in the larger particles. The quantitative differences in the blocking temperature as determined by the two different techniques are attributable to the differing sensitivities of the techniques. Specifically, the \Tb\ values determined by magnetometry are lower in temperature than the onset \Tb\ as detected by \muSR, which is consistent with the fact that \muSR\ is a faster probe. Fluctuations longer than $\sim 10^{-6}$~s begin to appear static to \muSR, so for a finite temperature range above \Tb\ as measured by magnetometry, the nanoparticle spins fluctuate faster than the measurement time scale for magnetometry but slower than that of \muSR. On the other hand, the midpoint of the blocking transition as measured by \muSR\ (corresponding to $f_{\mathrm{SPM}}=0.5$) occurs at lower temperatures than \Tb\ as measured by magnetometry, suggesting that \Tb\ as identified from the VSM data occurs when more than half of the sample volume is in the SPM state.

The greatest discrepancies between the magnetometry and \muSR\ results are for the magnetic reversal attempt time $\tau_0$. The values obtained from the VSM data using Eq.~\ref{eq:taustar} are orders of magnitude larger than expected. The inclusion of the Zeeman energy term in the analysis of the VSM data improves the estimate of $\tau_0$, but is still not sufficient to produce a physically reasonable value. This demonstrates the inherent challenges of extracting a reliable attempt time from bulk magnetometry data. On the other hand, the \muSR\ estimates of $\tau_0$ fall into the expected range of values and produce a self-consistent estimate of the blocking temperature, underscoring the accuracy of this approach. The favorable time scale of \muSR\ measurements, together with its sensitivity to the local magnetic environment, make this technique a powerful approach to probing SPM dynamics that can complement more commonly used experimental methods such as bulk magnetometry.

\begin{figure}
	\includegraphics[width=70mm]{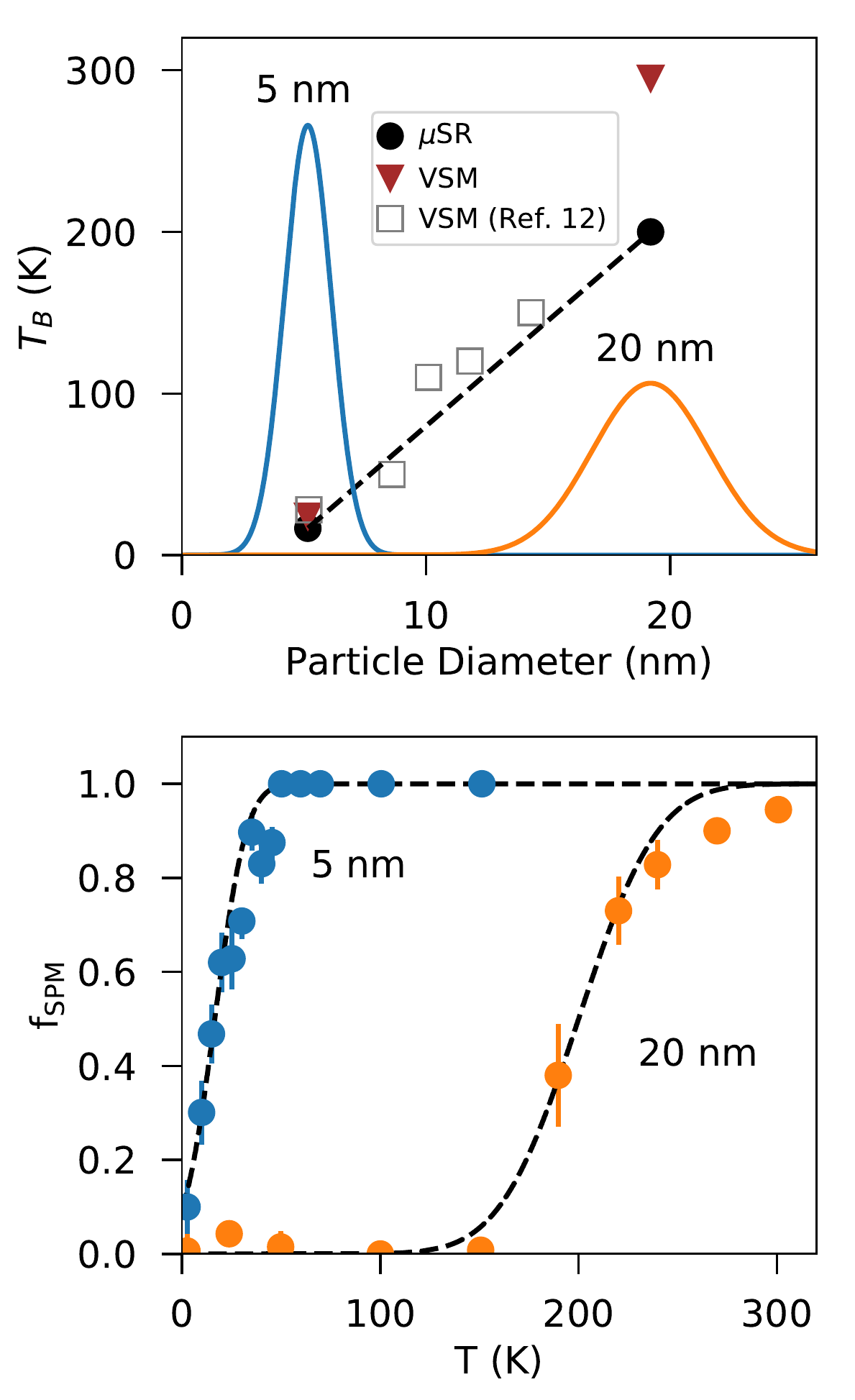}
	\caption{\label{fig:fracFits}(a) Blocking temperature \Tb\ versus particle size as determined by various techniques. In the case of the \muSR\ results (black circles), the value of \Tb\ shown here corresponds to the temperature at which 50\% of the sample is in the blocked state. The dashed line represents the equation of the line connecting the two \muSR\ data points. The solid blue and orange curves represent the particle size distributions for the two samples studied in this work. (b) Calculated temperature dependence of the superparamagnetic fraction $f_{\mathrm{SPM}}$ (dashed curves) compared to the experimentally determined results (solid circles) for the two samples in the present work. Details about the calculations are given in the main text.
	}		
\end{figure}

Finally, we now discuss in more detail the observation of a gradual transition from the blocked to the SPM state as revealed by the \muSR\ data. This indicates the presence of an intermediate mixed state containing coexisting SPM and blocked regions within the sample. The finite size distribution of each sample will be at least partially responsible for this gradual transition, since particles with slightly different sizes will enter the blocked state at slightly different temperatures. However, it is conceivable that other factors such as interparticle interactions could also contribute to the broadening of the blocking transition. To investigate this, we calculated the expected SPM fraction as a function of temperature for both samples assuming that \Tb\ depends linearly on particle size. This linear dependence was estimated by considering the midpoint blocking temperatures (where 50\% of the volume is blocked) for the two samples in the present work, and determining the equation of the line connecting those two points as a function of particle size. This is shown by the black circles and dashed line in Fig.~\ref{fig:fracFits}(a). The equation of the dashed line is $T_B(D) = AD+B$, where $D$ is the particle size in nm, $A = 13~\mathrm{K/nm},$ and $B = -50.8~\mathrm{K}$. For comparison, we additionally plot \Tb\ as determined from the ZFC peak in the magnetometry data as brown triangles and other previously published magnetometry results~\cite{klomp;ieeem20} as open squares. Other experimental stuides of magnetic NPs have likewise suggested a linear dependence of \Tb\ on particle size as a first order approximation in the 5 -25 nm range~\cite{mohap;cec13,shim;ssc08,aslan;ieeem18}. The particle size distributions $p(D)$ for the two samples in the present study are also shown by the blue and orange curves, approximating the measured size distributions for the 5 and 20~nm NPs displayed in Fig.~\ref{fig:5nmTEM-VSM}(a) and Fig.~\ref{fig:20nmTEM-VSM}(a), respectively, as normalized Gaussian distributions with the mean and standard deviation determined from the TEM characterization. 

The expected SPM fraction $f_{\mathrm{SPM}}$ can be calculated for a given temperature $T$ by inverting the linear function $T_B(D)$ to determine the cutoff particle size $D_c(T) = (T-B)/A$ that undergoes the blocking transition at $T$, and integrating the normalized particle size distribution $p(D)$ up to $D_c$. Mathematically, we have
\begin{align}\label{eq:frac}
f_{\mathrm{SPM}}(T) = \int_{0}^{D_c (T)}p(D)\mathrm{d}D.
\end{align}
The results of this calculation for both samples are shown as the dashed curves in Fig.~\ref{fig:fracFits}(b), together with the experimentally determined values shown as the colored circles. The agreement between the experimental and calculated results is relatively good, especially considering the simplicity of our model with an assumed linear dependence of \Tb\ on particle size. This indicates that the broad temperature range of coexisting blocked and superparmagnetic regions revealed by \muSR\ is primarily caused by the finite particle size distribution, although interparticle interactions or other factors may play a role as well~\cite{bruve;jap15,majet;jpd06}.

In summary, we have employed magnetometry and \muSR\ in a complementary fashion to gain detailed information about the SPM behavior and the blocking transition in assemblies of \feo\ nanoparticles with average particle diameters of 5~nm and 20~nm. We established that the transition between the SPM and blocked states extends gradually throughout the sample volume over a wide temperature interval due to the finite size distribution of each sample. We provided estimates of the blocking temperature for each sample using various methods, and we extracted microscopic information including the activation energy, anisotropy constant, and magnetic reversal attempt time. This allowed us to investigate the dependence of these properties on particle size. Finally, we have also discussed some of the difficulties of estimating microscopic parameters such as the attempt time $\tau_0$ using bulk magnetometry alone, and instead have shown that local probes such as \muSR\ can be highly useful for reliably estimating $\tau_0$, determining the volume fraction of competing states, and otherwise complementing data obtained from bulk probes.

\textbf{Acknowledgements}
	
We thank Gerald Morris, Bassam Hitti, and Donald Arseneau for their support at the \muSR\ beamline at TRIUMF. This work was made possible by funds provided by the College of Physical and Mathematical Sciences at Brigham Young University.

%

\end{document}